\documentstyle[11pt]{article}
\newcommand{\beq}{\begin{equation}}
\newcommand{\eeq}{\end{equation}}
\newcommand{\beqa}{\begin{eqnarray}}
\newcommand{\eeqa}{\end{eqnarray}}

\title{Entanglement-Enhanced Classical Communication
on a Noisy Quantum Channel}
\author{\protect Charles H. Bennett$^{(1)}$, Christopher A.
Fuchs$^{(2,3)}$, and John A. Smolin$^{(1)}$\medskip \\
$^{(1)}$ IBM Research Division, Yorktown Heights, NY 10598\\ 
$^{(2)}$ D\'epartement IRO, Universit\'e de Montr\'eal, C.~P.
6128, Succursale centre-ville,\\ Montr\'eal, Canada H3C 3J7\\
$^{(3)}$ Present address: Norman Bridge Laboratory of Physics 12-33,
California\\ Institute of Technology, Pasadena, CA 91125}
\date{29 October 1996}

\begin{document}

\maketitle

\begin{abstract}
We consider the problem of trying to send a single classical bit
through a noisy quantum channel when two transmissions through
the channel are available as a resource.  Classically, two
transmissions add nothing to the receiver's capability of inferring
the bit.  In the quantum world, however, one has the possible further 
advantage of entangling the two transmissions.  We demonstrate
that, for certain noisy channels, such entangled transmissions
enhance the receiver's capability of a correct inference.
\end{abstract}

\section{Introduction}

Much of the growing field of quantum information theory is founded
upon the asking of a single question: for a given task, can it
help to use an entangled pair of particles as a resource?  In this
contribution, we ask just this question again, but in the context
of a simple classical communication problem.  The task is to use a
noisy quantum channel as effectively as possible for the
communication of a {\it single\/} bit, 0 or 1, given that only
two transmissions through the channel are 
allowed.\footnote{Information theoretically, this means that Alice
sends each message with equal probability. The methods described here
apply, with slight modification, to the case of unequal probabilities
as well.  We focus on the case of a full bit because the significant
quantum behavior reported herein can be illustrated without such
complications.}  Alice, the
transmitter, may use any means allowed by the laws of quantum
mechanics for encoding the bit in the two transmissions; Bob, the 
receiver, may use any means allowed by quantum mechanics for measuring
the output of the channel in an attempt to infer the actual bit.
The question is, can
the use of an encoding that takes advantage of entanglement
between the transmissions increase Bob's capability of inference?

As an example, the channel in question could be a somewhat
depolarizing fiber optic cable. The two transmissions would then be 
represented by two separate photons being sent through the cable.
In this case, the bit is to be encoded somehow in the polarization 
degrees of freedom of the photons, each represented by a
two-dimensional Hilbert space ${\cal H}_2$.  For this, our question
boils down to whether the optimal encoding for such a channel will be
in terms of product states---say by some 
$|\psi_0\rangle|\psi_0\rangle$ and $|\psi_1\rangle|\psi_1\rangle$
for 0 and 1 respectively---or rather by some {\it entangled\/} states
$|\Psi_0\rangle$ and $|\Psi_1\rangle$ on
${\cal H}_2\otimes{\cal H}_2$.

Classically, any attempt to protect the bit from noise by redundancy
is always stalemated when only two transmissions are available.  This 
is because Bob cannot perform any sort of ``majority vote'' error
correction on two transmissions.  In the quantum version of the
problem, as posed above, however, things become much more interesting.
Extensive numerical work shows that there are indeed noise models for
which entangled transmissions are more reliable for carrying the
bit from sender to receiver than would be the case otherwise.  In
fact, such noise models seem to be the rule rather than the exception.

This paper is devoted to demonstrating the existence of this
effect---i.e., entanglement-enhanced classical communication---for
a particularly simple noisy quantum channel, the ``two-Pauli
channel.''  In Section 2 below, we develop the formalism required to 
give the general problem a precise statement.  In Section 3, we
introduce the two-Pauli channel and analyze what can be done with it
upon one transmission and, alternatively, upon two transmissions but
with product-state inputs.  In Section 4, we find the optimal 
(entangled) encoding for two transmissions through the channel.  We
close the paper in Section 5 with a brief discussion of another
channel of interest.  Also we pose the deeper information-theoretic
question of whether entanglement can be used to increase the 
classical information carrying capacity of a noisy quantum channel.

\section{Noisy Channel Preliminaries}

The general description of the problem is the following.  The action
of a noisy channel on the physical system transmitted through it
is described by a mapping $\phi$ from input density operators to 
output density operators on that system.  In all cases, the mapping
is assumed to come about via an interaction between the system of 
interest and an independently prepared environment, to which neither 
Alice nor Bob has access.  Thus a noisy channel is captured 
formally by a mapping of the form
\begin{equation}
\rho\;\longrightarrow\;\phi(\rho)={\rm tr}_{\scriptscriptstyle\rm E}
\Big(U(\rho\otimes\sigma)U^\dagger\Big)\;,
\label{Naugahide}
\end{equation}
where $\sigma$ is the initial state of the environment, $U$ is
the unitary interaction between the system and environment, and
${\rm tr}_{\scriptscriptstyle\rm E}$ denotes a partial trace over the
environment's Hilbert space.  For specificity, we assume $\rho$ is
a density operator on a $d$-dimensional Hilbert space ${\cal H}_d$.

A convenient means of representing all possible noise models is
given by the Kraus representation theorem \cite{Kraus83,Choi75}.  This
states that Eq.~(\ref{Naugahide}) can be written in the form
\begin{equation}
\phi(\rho)=\sum_i A_i\rho A_i^\dagger\;,
\label{AnklesOfHair}
\end{equation}
where the $A_i$ satisfy the completeness relation
\begin{equation}
\sum_i A_i^\dagger A_i = I\;.
\label{HankThoreau}
\end{equation}
Conversely, any set of operators $A_i$ satisfying 
Eq.~(\ref{HankThoreau}) can be used in Eq.~(\ref{AnklesOfHair})
to give rise to a valid noisy channel in the sense of 
Eq.~(\ref{Naugahide}).

In the language of the Kraus theorem, the situation
of two transmissions through the channel is described by the mapping
\begin{equation}
R\;\longrightarrow\;\Phi(R)=\sum_{i,j}
(A_i\otimes A_j)R(A_i\otimes A_j)^\dagger\;,
\label{Babette}
\end{equation}
where $R$ denotes a density operator on the $d^2$-dimensional
Hilbert space ${\cal H}_d\otimes{\cal H}_d$.

In our particular problem,
Alice encodes the bit she wishes to transmit to Bob by preparing a 
quantum system in one of two states $R_0$ or $R_1$
on ${\cal H}_d\otimes{\cal H}_d$.  The action of the channel via
Eq.~(\ref{Babette}) leads to
one of two possible output density operators for Bob at the receiving 
end, say either $\tilde{R}_0$ or $\tilde{R}_1$ respectively.

For the problem of distinguishing the two density operators 
$\tilde{R}_0$ and $\tilde{R}_1$, we imagine Bob performing a
general quantum mechanical measurement, or positive operator-valued
measure (POVM) $\{E_b\}$, and then using the acquired data to
venture a guess about the identity of the density operator.  Depending
upon which bit $s$ Alice has sent, the probability of Bob's 
measurement outcomes will be given by ${\rm tr}(\tilde{R}_s E_b)$.
Clearly the best strategy for Bob in identifying the bit---upon
finding an outcome $b$---is to guess the value of $s$ for which
${\rm tr}(\tilde{R}_s E_b)$ is the largest.  Since each input is
equally likely, this gives rise to an average probability of error 
given by
\beq
P_e\Big(\{E_b\}\Big)=\frac{1}{2}\sum_b\min\Big\{ {\rm tr}
(\tilde{R}_0 E_b),{\rm tr}(\tilde{R}_1 E_b)\Big\}\;.
\label{SelfReference}
\eeq
This makes it clear that the best measurement on Bob's part is to
choose the one that minimizes this expression.  It turns out that
this measurement can be described by a standard von~Neumann
measurement of the Hermitian operator
$\Gamma = \tilde{R}_1-\tilde{R}_0$ \cite{Helstrom76}.  Moreover, with
this measurement, Eq.~(\ref{SelfReference}) reduces to \cite{Fuchs96}
\beq
P_e = \frac{1}{2} - \frac{1}{4}{\rm tr}\Big|\tilde{R}_1-\tilde{R}_0
\Big|\;.
\label{BigMan}
\eeq
Here ${\rm tr}|A|$, for any Hermitian operator $A$, should be
interpreted as the sum of the absolute value of $A$'s eigenvalues.

The question we ask in this paper can now be stated in a precise
manner.  For any two orthogonal pure state inputs $|0\rangle$ and 
$|1\rangle$ on ${\cal H}_d\otimes{\cal H}_d$,
what is the smallest possible value that
\beqa
P_e
&=&
\frac{1}{2}-\frac{1}{4}{\rm tr}\Big|\,\Phi(|1\rangle\langle 1|)-
\Phi(|0\rangle\langle 0|)\,\Big|
\nonumber\\
&=&
\frac{1}{2}-\frac{1}{4}{\rm tr}\Big|\,\Phi\Big(|1\rangle\langle 1|-
|0\rangle\langle 0|\Big)\,\Big|
\rule{0mm}{7mm}
\label{RubyNell}
\eeqa
can take?  And, more importantly, is it ever the case that the
smallest value can be achieved only by entangled states?\footnote{Of
course, in principle, one must consider the possibility that the
optimal inputs might be nonorthogonal or even mixed states.  A proof
that such possibilities are less than optimal is given in the 
Appendix.  This is in accord with the intuition that inputs with
less than maximal distinguishability cannot help for this problem.}

\section{The Two-Pauli Channel}

The {\it two-Pauli channel\/} is a noisy quantum channel on a single
qubit, ${\cal H}_2$, described by three Kraussian
$A_i$ operators:
\begin{equation}
A_1=\sqrt{x}\,I\;,\;\;\;\;\;
A_2=
\sqrt{ {\scriptstyle \frac{1}{2} }
(1-x)}\,\sigma_1\;,\;\;\;\;\;
A_3=-i\sqrt{ {\scriptstyle \frac{1}{2} }
(1-x)}\,\sigma_2\;,
\label{MamaMan}
\end{equation}
where $I$ is the identity operator and $\sigma_1$, $\sigma_2$, and 
$\sigma_3$ are the standard Pauli matrices, i.e.,
\beq
\sigma_1=\left(\begin{array}{cr}
0 & 1\\
1 & 0\end{array}\right),\;\;\;\;
\sigma_2=\left(\begin{array}{cr}
0 & -i\\
i & 0\end{array}\right),\;\;\;\;
\sigma_3=\left(\begin{array}{cr}
1 & 0\\
0 & -1\end{array}\right).
\eeq
This channel has a simple interpretation:  with probability
$x$, it leaves the qubit alone; with probability $1-x$ it randomly
applies one of the two Pauli rotations to the qubit.

Note that, because there is no $A_i$ in Eq.~(\ref{MamaMan})
corresponding to the Pauli matrix $\sigma_3$, the two-Pauli channel
cannot be thought of as a simple depolarizing channel.  Nor can it
be thought of as a simple dephasing channel, where only one
Pauli rotation $\sigma_j$ acts on the qubit.  A classical bit can be 
sent perfectly through a dephasing channel by choosing $|0\rangle$
and $|1\rangle$ to be eigenstates
of $\sigma_j$.  Moreover, numerical work demonstrates that there is
no benefit from using entangled transmissions for the depolarizing 
channel.  It turns out that the asymmetry of the two-Pauli channel is 
just right for seeing the entanglement enhancement effect in
a particularly clean way.
  
Let us gain some intuition about the two-Pauli channel by
first considering only one transmission through it.
Suppose we consider two possible (commuting) inputs, $\rho_+$ and
$\rho_-$, given in Bloch sphere representation by
\beq
\rho_\pm=\frac{1}{2}\Big(I\pm\vec{a}\cdot\vec{\sigma}\Big)\;.
\eeq
Here, $\vec{a}=(a_1,a_2,a_3)$ is any real vector of length 1 or less,
and $\vec{\sigma}$ is the vector of Pauli matrices.

One can easily verify that the action of the channel on these
two density operators is:
\beq
\rho_\pm\;\longrightarrow\;\phi(\rho_\pm)=
\frac{1}{2}\Big(I\pm\vec{b}\cdot\vec{\sigma}\Big)\;,
\eeq
where
\beq
\vec{b}=\Big(a_1 x,\,a_2 x,\,a_3 (2x-1)\Big)\;.
\eeq
Note that one transmission through the channel takes commuting
density operators to commuting density operators.  We shall see that
this feature is not necessarily true for two transmissions.

If we now specifically consider the channel for communication
purposes, we should make the final states as distinguishable as
possible.  This means we should pick the vector $\vec{a}$ so that
$\phi(\rho_+)$ and $\phi(\rho_-)$ are as pure as possible.  How to
do this will depend upon the value of the parameter $x$, since the
eigenvalues of both $\rho_+$ and $\rho_-$ are
\beq
\frac{1}{2}\pm\frac{1}{2}\sqrt{(a_1^2+a_2^2)x^2+a_3^2(2x-1)^2}\;.
\eeq
If $x\ge\frac{1}{3}$, then $|x|\ge|2x-1|$.  Hence, clearly, the
optimal inputs will be pure states with $a_3=0$.  If, on the other
hand, $x\le\frac{1}{3}$, then $|x|\le|2x-1|$.  The optimal inputs in
this case will again be pure states but with $a_1=a_2=0$.  

The probability of error in guessing the identity of the states
works out easily enough.  It is just
\beq
P_e = \left\{\begin{array}{ll}
x & \;\;\mbox{if}\:\;\;\;\; x\le\frac{1}{3}\\
\frac{1}{2} - \frac{1}{2}x& \;\;\mbox{if}\:\;\;\;\; x\ge\frac{1}{3}
\rule{0mm}{5mm}\end{array}\right]\;.
\label{ChocolateBar}
\eeq

With this much of an introduction to the two-Pauli channel, let us
now briefly consider two transmissions through the channel, but with
those transmissions restricted to be product states.  There is
not much to be said here.  If we assume two inputs of the form
\beq
R_0=\rho_0\otimes\rho_0\;\;\;\;\;\;\mbox{and}\;\;\;\;\;\;
R_1=\rho_1\otimes\rho_1\;,
\label{CurlyMoe}
\eeq
then the error probability remains the same as above.  This can be
corroborated easily both from classical principles and quantum
principles.  Since $\tilde{R}_0$ and $\tilde{R}_1$ themselves commute
when the inputs are orthogonal and pure, working out
Eq.~(\ref{BigMan}) in this case is hardly more difficult than for
the single transmission case.  The best possible error probability
remains that listed in Eq..~(\ref{ChocolateBar}).

\section{Entangled Transmissions}

In this Section, we turn to the problem of sending entangled
transmissions down the channel.  
A convenient basis with which to write the input states for two
transmissions through the two-Pauli channel is the Bell operator
basis, where the four orthonormal basis vectors are
\begin{equation}
|\Phi^\pm\rangle=\frac{1}{\sqrt{2}}\Big(|\uparrow\,\rangle|\uparrow\,
\rangle\pm|\downarrow\,\rangle|\downarrow\,\rangle\Big)
\;\;\;\;\;\;\mbox{and}\;\;\;\;\;\;
|\Psi^\pm\rangle=\frac{1}{\sqrt{2}}\Big(|\uparrow\,\rangle|\downarrow
\,\rangle\pm|\downarrow\,\rangle|\uparrow\,\rangle\Big)\;.
\end{equation}
We use the usual notation that
\beq
|\uparrow\,\rangle\;=\left(\begin{array}{c} 1 \\ 0\end{array}\right)
\;\;\;\;\;\;\mbox{and}\;\;\;\;\;\;
|\downarrow\,\rangle\;=\left(\begin{array}{c} 0 \\ 
1\end{array}\right)
\label{BrendaAndBelinda}
\eeq
for the eigenvectors of $\sigma_3$.
With respect to this basis, an arbitrary set of two input states 
can be written as
\beq
|0\rangle = a_0 |\Phi^+\rangle + b_0 |\Phi^-\rangle + c_0
|\Psi^+\rangle + d_0 |\Psi^-\rangle
\eeq
and
\beq
|1\rangle = a_1 |\Phi^+\rangle + b_1 |\Phi^-\rangle + c_1
|\Psi^+\rangle + d_1 |\Psi^-\rangle\;.
\eeq

Making the effort to work through Eq.~(\ref{Babette}), we
find that the output density operator
\beq
\tilde{R}_0=\Phi(|0\rangle\langle0|)
\eeq
expressed in the basis of Eq.~(\ref{BrendaAndBelinda}) is
$$
\left[\begin{array}{cccc}
e a_0^2 + f b_0^2 + g c_0^2 + g d_0^2 & h a_0 b_0 &
x a_0 c_0 & k a_0 d_0 \\
h a_0 b_0 & f a_0^2 + e b_0^2 + g c_0^2 + g d_0^2 &
k b_0 c_0 & x b_0 d_0
\rule{0mm}{6mm}\\
x a_0 c_0 & k b_0 c_0 &
g a_0^2 + g b_0^2 + e c_0^2 + f d_0^2 & h c_0 d_0
\rule{0mm}{6mm}\\
k a_0 d_0 & x b_0 d_0 & h c_0 d_0 &
g a_0^2 + g b_0^2 + f c_0^2 + e d_0^2
\rule{0mm}{6mm}
\end{array}\right]
$$
where
\beqa
e &=& \frac{1}{2}(1-2x+3x^2) \\
f &=& \frac{1}{2}(1-x)^2
\rule{0mm}{6mm}\\
g &=& x(1-x)
\rule{0mm}{6mm}\\
h &=& 2x-1
\rule{0mm}{6mm}\\
k &=& x(2x-1)
\rule{0mm}{6mm}
\eeqa
A similar expression holds for $\Phi(|1\rangle\langle1|)$ but with
1 exchanged for 0 everywhere.

Before tackling the problem at hand, it is worthwhile exploring a
few features of this noise model.  For instance, it would be
convenient if it worked out that, as with the product
state, whenever $|0\rangle$ and $|1\rangle$ are orthogonal,
$\Phi(|0\rangle\langle0|)$ and $\Phi(|1\rangle\langle1|)$ were
assured to commute.  This, unfortunately, is not the case.  A
simple counterexample suffices to show this.  Simply take
\beq
|0\rangle = \left(\begin{array}{r}
-0.459506 \\ -0.870791 \\ 0.127295 \\ 0.119889\end{array}\right)
\;\;\;\;\;\;\mbox{and}\;\;\;\;\;\;
|1\rangle = \left(\begin{array}{r}
-0.578111 \\ 0.163069 \\ -0.770549 \\ -0.213192\end{array}\right)\;.
\eeq
Then $\langle0|1\rangle=0$, but
\beq
\Big[\Phi(|1\rangle\langle 1|),\;\Phi(|0\rangle\langle 0|)\Big]
\ne0\;.
\eeq
Interestingly enough, however, there are special cases where the
commutativity of the two outputs is assured.  For instance,
take
\beqa
|0\rangle &=& \cos\alpha\,|B_1\rangle+\sin\alpha\,|B_2\rangle\\
|1\rangle &=& -\sin\alpha\,|B_1\rangle+\cos\alpha\,|B_2\rangle\;,
\rule{0mm}{6mm}
\eeqa
where $|B_1\rangle$ and $|B_2\rangle$ are any two Bell states.  Then
it is easily checked that $\Phi(|0\rangle\langle0|)$ and
$\Phi(|1\rangle\langle1|)$ do indeed commute in this case.

As an alternate example, let $|0\rangle$ be any non-Bell state
vector in the plane spanned by $|\Phi^+\rangle$ and $|\Psi^+\rangle$, 
and let $|1\rangle$ be any non-Bell state vector in the plane
spanned by $|\Phi^-\rangle$ and $|\Psi^-\rangle$.
It turns out that the outputs of the two-Pauli channel due
to these inputs never commute {\it except\/} in the case that the
channel parameter $x$ equals either $0$, $1$, or $1/3$.

These examples give some hint that the two-Pauli channel is channel
fairly rich in structure.  So, with this,
let us return to the question of the optimal input states for
two transmissions.  Numerical work demonstrates that for channel 
parameter values $x\le1/3$, the optimal inputs are product states
of the form given by Eq.~(\ref{CurlyMoe}).  However, for channel 
parameters $1/3<x<1$, entangled inputs give the minimal error 
probabilities.  Moreover, within the latter regime, though
there appear to be many equivalent optimal entangled signals, 
two inputs can always be taken to be of the form
\begin{equation}
|0\rangle=\cos\alpha\,|\Phi^+\rangle+\sin\alpha\,|\Psi^+\rangle,
\label{Moe}
\end{equation}
and
\begin{equation}
|1\rangle=-\sin\alpha\,|\Phi^+\rangle+\cos\alpha\,|\Psi^+\rangle
\label{Larry}
\end{equation}
without any loss of performance.

The remainder of this Section is devoted to fleshing out the 
consequences of taking Eqs.~(\ref{Moe}) and (\ref{Larry}) as an 
ansatz in our problem.  The best probability of error in Bob's
inference of the signal, in accordance with Eq.~(\ref{RubyNell}),
follows after some algebra:
\beq
P_e(\alpha)\,=\,
\frac{1}{2}\,-\,\frac{1}{2}\left(
\left(\frac{1}{4}\Big(1-4x+5x^2\Big)^2\cos^2 2\alpha +x^2\sin^2 
2\alpha\right)^{\!1/2}\,+\,
\frac{1}{2}(1-x)\Big|(1-3x)\cos 2\alpha\Big|\right).
\label{Wyatt}
\eeq
If we define
\beqa
F &=& \frac{1}{4}(1-x)(1-5x)(1-2x+5x^2) \\ 
G &=& \frac{1}{2}(1-x)(1-3x)
\rule{0mm}{6mm}\\
Z &=& \cos 2\alpha \;,
\label{Ignacio}
\rule{0mm}{6mm}
\eeqa
the error probability as a function of the ansatz can be written more 
compactly as
\beq
P_e(Z) = \frac{1}{2}-\frac{1}{2}\left(\sqrt{F Z^2+x^2}+|GZ|\right)\;.
\label{Audrey}
\eeq

Our task now reduces to optimizing the ansatz in order to find the
two inputs that lead to the most distinguishable outputs.  This is
done by extremizing Eq.~(\ref{Audrey}):
\begin{equation}
\frac{\partial P_e(Z)}{\partial Z}=0\;.
\end{equation}
This variational equation will have a solution for an optimal
$Z$ as long as $x$ is such that $Z^2$ remains within the range set 
by Eq.~(\ref{Ignacio}), i.e., between 0 and 1. 
This occurs for
\beq
0\le\frac{G^2 x^2}{F(F-G)}\le1\;,
\eeq
which, in turn, requires that
\beq
x\;\ge\;\frac{4}{15} - \frac{41}{30}\Big(15\sqrt{330}-73\Big)^{-1/3}
+ \frac{1}{30}\Big(15\sqrt{330}-73\Big)^{1/3}\;\approx\;
0.227539 \rule{0mm}{6mm}
\eeq
Since we only need solutions for $x\ge1/3$, this implies that
our ansatz at least remains valid within the range of interest.

For a given $x$, the optimal $Z^2$ works out to be given by
\begin{equation}
Z^2 =\frac{(1-3x)^2}{4x(5x-1)(1-2x+5x^2)}.
\end{equation}
Hence the optimal version of Eq.~(\ref{Wyatt}) reduces to
\beq
P_e = \frac{1}{2} - 2\sqrt{\frac{x^5}{(5x-1)(1-2x+5x^2)}}\;.
\eeq

This demonstrates our point:  as long as $1/3<x<1$, entangled
transmissions through the two-Pauli channel are more effective
at disabling noise than product state transmissions.  To convey a 
feeling for the effectiveness of entangled transmissions, 
we tabulate a few representative points below.

\begin{center}
\begin{tabular}{|c|c|c|} \hline
x   & $P_e$ (product states) & $P_e$ (entangled states) \\
\hline\hline
.50  & 0.250000               & 0.241801 \\
.60  & 0.200000               & 0.188231 \\
.70  & 0.150000               & 0.137817 \\
.80  & 0.010000               & 0.090072 \\
.90  & 0.050000               & 0.044319 \\
.95  & 0.025000               & 0.022009 \\ \hline
\end{tabular}
\end{center}

\section{Discussion}

This paper has been largely devoted to developing a formal framework
for tackling the question of entanglement-enhanced classical
communication and demonstrating the existence of this effect for a
particular noisy channel, the two-Pauli channel.  However, computer 
simulations further corroborate that this example is not in any way 
isolated: it may be a property of most noisy channels.  For instance,
another example where entangled transmissions are effective is a
``amplitude damping channel,'' where the qubit
arises from either one or no photons in a mode.  The noise in this
channel is due to the possibility of a photon leaking off to 
infinity.  The Kraussian $A_i$ operators for this
channel \cite{Chuang97} are
\beq
A_1=\left(\begin{array}{cc}
\sqrt{x} & 0 \\
0 & 1
\end{array}\right)\;\;\;\;\;\;\mbox{and}\;\;\;\;\;\;
A_2=\left(\begin{array}{cc}
0 & 0 \\
\sqrt{1-x} & 0
\end{array}\right)\;.
\eeq
and an analysis similar to the preceding one for the two-Pauli
channel can be carried out in like manner.

Finally, let us emphasize that what we have shown here is that, by 
increasing our resources from one to two transmissions and allowing
those transmissions to be entangled, we can make a classical bit more
resilient to noise.  This has no analog in classical information 
theory.  A deeper question arises from comparing the increase in
resources to the overall information transmittable by those 
resources.  This is the question of whether the classical information 
capacity of a noisy quantum channel can be increased by entangling 
transmissions \cite{Bennett96}.  Let us close the paper by making
this question precise.

If $n$ possible inputs to a channel---used with prior probabilities
$\pi_1$, $\pi_2$, \ldots, $\pi_n$---lead to $n$ distinct density 
operators $\rho_1$, $\rho_2$, \ldots, $\rho_n$ at the output (with
like prior probabilities), then for a fixed POVM $\{E_b\}$, the 
mutual information recoverable about the identity of the input is
\beq
I\Big(\{\pi_i\},\{E_b\}\Big)=-\sum_b{\rm tr}(\rho E_b)\log
{\rm tr}(\rho E_b)+
\sum_{i=1}^n\pi_i\sum_b{\rm tr}(\rho_i E_b)\log{\rm tr}(\rho_i E_b)\;,
\eeq
where
\beq
\rho=\sum_{i=1}^n\pi_i\rho_i\;.
\eeq
This is the Shannon information of the output symbols minus the 
average Shannon information of the output symbols conditioned on the 
input.  The channel capacity for the given set of inputs is found by
optimizing the prior probabilities of the inputs and optimizing the
quantum measurement used at the output:
\beq
C(\rho_1,\,\rho_2,\,\ldots,\,\rho_n)=\max_{\{\pi_i\}}\,\max_{\{E_b\}}
I\Big(\{\pi_i\},\{E_b\}\Big)\;.
\eeq
This defines the ultimate information carrying capacity for a single
transmission through the channel as a function of the particular 
input quantum states.

What we are in search of is a comparison of the best possible
capacity (as a function of the inputs) for one transmission versus
the same for two transmissions.  That is to say, under completely 
general uses of the channel, what is the best possible channel 
capacity
\beq
C\Big(\Phi(|1\rangle\langle 1|),\,\Phi(|2\rangle\langle 2|),\,\ldots,
\,\Phi(|n\rangle\langle n|)\Big)\;,
\label{WearyKnees}
\eeq
where an optimization must also encompass the number of inputs $n$?
Moreover, how does this number compare to a similarly defined
capacity $C_1$ for a single transmission through the channel?  If
Eq.~(\ref{WearyKnees}) turns out to be more than twice $C_1$, then
we would have that classical information capacities are
``super-additive'' with respect to multiple uses of the same channel.
The existence of this phenomena would be yet another surprising 
quantum effect indeed.

\section{Appendix: Why Orthogonal Signal States?}

Suppose we have any quantum channel whatsoever and that its action
on density operators is given (in standard form) by a
trace-preserving completely positive map
\beq
\rho\;\longrightarrow\;{\cal E}(\rho)=\sum_i B_i\rho B_i^\dagger\;,
\eeq
where
\beq
\sum_i B_i^\dagger B_i = I\;.
\eeq
For the problem of finding the two inputs that lead to two maximally 
distinguishable outputs (as in these notes), how do we know that the 
optimal inputs should pure states rather than mixed?  Given that the 
inputs are pure states, how do we know that the optimal ones must be
orthogonal?

If we grant two standard facts from linear algebra
\cite{Marshall79}, then we can answer both these questions quite 
readily.  The first fact is that, 
for any operator $A$,
\beq
\max_U\;{\rm Re}\;{\rm tr}(UA)={\rm tr}|A|\;
\eeq
where the maximum is taken over all unitary operators $U$.  The second
fact is that, for any two $n\times n$ Hermitian operators $A$ and $B$,
\beq
\sum_{i=1}^n\lambda_{n-i+1}(A)\lambda_i(B)\le{\rm tr}(AB)\le
\sum_{i=1}^n\lambda_i(A)\lambda_i(B)\;,
\label{VeraLynn}
\eeq
where $\lambda_i(X)$ denotes the $i$'th eigenvalue of $X$ when
enumerated in nonincreasing order.

Now, in order to minimize the probability of error in carrying one
bit across the given channel, we must find two states $\rho_1$ and
$\rho_0$ such that
\beq
{\rm tr}\Big|{\cal E}(\rho_1)-{\cal E}(\rho_0)\Big|
\eeq
is maximized.  Therefore, let us focus on this expression.  Define
the ``conjugate'' mapping ${\cal E}^*$ to ${\cal E}$ by the action
\beq
X\;\longrightarrow\;{\cal E}^*(X)=\sum_i B_i^\dagger X B_i\;.
\eeq
Then
\beqa
{\rm tr}\Big|{\cal E}(\rho_1)-{\cal E}(\rho_0)\Big|
&=&
{\rm tr}\Big|{\cal E}(\rho_1-\rho_0)\Big|
\nonumber\\
&=&
\max_U\;{\rm Re}\;{\rm tr}\Big(U{\cal E}(\rho_1-\rho_0)\Big)
\nonumber\\
&=&
\max_U\;\frac{1}{2}{\rm tr}\Big((U+U^\dagger)
{\cal E}(\rho_1-\rho_0)\Big)
\nonumber\\
&=&
\max_U\;\frac{1}{2}{\rm tr}\Big((\rho_1-\rho_0){\cal E}^*(U+U^\dagger)
\Big)
\nonumber\\
&=&
\max_U\;\frac{1}{2}\left[
{\rm tr}\Big(\rho_1{\cal E}^*(U+U^\dagger)\Big) -
{\rm tr}\Big(\rho_0{\cal E}^*(U+U^\dagger)\Big) \right]\;.
\eeqa
(For the second to last step in this, we used the cyclic property of
the trace.)  Note that, because $U+U^\dagger$ is an Hermitian 
operator, the operator ${\cal E}^*(U+U^\dagger)$ is also Hermitian.

Let us focus, for the moment, on any particular unitary operator $U$
in the maximization procedure above.  Using Eq.~(\ref{VeraLynn}), we
have
\beq
{\rm tr}\Big(\rho_1{\cal E}^*(U+U^\dagger)\Big)\;\le\;
\sum_{i=1}^n\lambda_i(\rho_1)
\lambda_i\Big({\cal E}^*(U+U^\dagger)\Big)\;\le\;
\lambda_1\Big({\cal E}^*(U+U^\dagger)\Big)
\eeq
and
\beq
-{\rm tr}\Big(\rho_0{\cal E}^*(U+U^\dagger)\Big)\;\le\;
-\sum_{i=1}^n\lambda_i(\rho_0)
\lambda_{n-i+1}\Big({\cal E}^*(U+U^\dagger)\Big)\;\le\;
-\lambda_n\Big({\cal E}^*(U+U^\dagger)\Big)\;.
\eeq
However, if $\rho_1$ and $\rho_0$ are chosen to be the eigenprojectors
of ${\cal E}^*(U+U^\dagger)$, then equality will be achieved
throughout these equations.  Thus for any particular $U$,
\beq
\max_{\rho_0,\rho_1}\;
{\rm tr}\Big((\rho_1-\rho_0){\cal E}^*(U+U^\dagger)\Big)=
\lambda_1\Big({\cal E}^*(U+U^\dagger)\Big)
-\lambda_n\Big({\cal E}^*(U+U^\dagger)\Big)
\eeq
and $\rho_1$ and $\rho_0$ must be orthogonal pure states to achieve
this.

Therefore, it follows that the input states optimal for leading to
the maximal distinguishability of the associated outputs will be
orthogonal pure states.

\section{Acknowledgments}
We would like to thank David DiVincenzo and Bill Wootters for
discussions.  C. A. F. was supported by NSERC.

\end{document}